\documentclass{article}

\usepackage{arxiv}

\usepackage[utf8]{inputenc} % allow utf-8 input
\usepackage[T1]{fontenc}    % use 8-bit T1 fonts
\usepackage{caption}
\usepackage{subcaption}
\usepackage{amsmath}
\usepackage{booktabs}       % professional-quality tables
\usepackage{amsfonts}       % blackboard math symbols
\usepackage{nicefrac}       % compact symbols for 1/2, etc.
\usepackage{microtype}      % microtypography
\usepackage{lipsum}
\usepackage{graphicx}
\graphicspath{ {./images/} }

\title{On the Computational Complexity of Determining Value}

\author{
 Tiasa Mondol, MMath \\
  \texttt{tiasa.ap.10@gmail.com}
   
}

\begin{document}
\maketitle
\begin{abstract}
\begin{quote}
    We look at an algorithmic information theory based definition of value of a creative artifact and discuss the computational difficulty associated with the creation or determination of value. We look at the computational resources required to create a valuable product and discuss how likely an observer is to verify a claim that an object required significant effort on the creator's part.
\end{quote} 
\end{abstract}

% keywords can be removed
%\keywords{First keyword \and Second keyword \and More}

\section{Introduction}
We start by building an algorithmic information theory based definition of value of a creative artifact. An artifact can be described concisely by an algorithm that generates a binary description of it on a computer (such formulation may seem reductive but it makes analysis of its complexity, the creative process and effort that went behind it much easier). We fix this particular computer $C$ as the only way to generate a description of an artifact which upon receiving some program $p$ (a set of instructions) halts after outputting only our desired object $x$ and nothing else. Algorithmic information theory says, in this settings, the length of shortest program is the object's Kolmogorov complexity \cite{b1} which is its absolute information content (a concise set of instructions for composing Bach's Cello Suites may be more economical and certainly more informative than the verbatim compositions themselves).
$$K_C(x)=\min_p\{|p|:C(p)=x\}$$
If the input programs are binary and we generate them via coin tosses, then the probability that $x$ will be the output on any given input is $Q_C(x)=\sum_{p:C(p)=x}2^{-|p|}$. This is remarkable as it is also the algorithmic probability of $x$'s existence on $C$. For the sum to be a proper probability distribution such that $Q_C(x)\leq 1$, it is however required that $C$ knows when to stop reading the input (due to Kraft inequality \cite{b1}) and such property is known as prefix-free where no program is a prefix of another valid program. Then the shortest program $p$ contributes the most to $x$'s probability and it is $p$'s computational effort to create $x$ that will determine $x$'s value.
 \section{Value as Computational Effort}
 Note that this ``effort'' can measured be in required time, space or energy; all of which are resources that we may have restricted access to and thus will naturally value an object to create which an extraordinary amount of these resources was needed. However, we continue our discussion focusing only on time as the other resources increase proportionally with it after a certain threshold (a computer which embarks on a long computation generates heat when it manipulates or erases information \cite{a125}; to cool it down we need to spend energy still. The discussion on assigning value to space is rather tricky; because space can be reused and even the polynomial space complexity class $PSPACE$ contains some of the most difficult problems, it is more interesting to discuss problems with bounded time). We define here a time-bounded version $K^t_C(x)$ which is the length of shortest program $p$ that outputs $x$ within a computable time bound $t(|x|)$.
 \begin{align*}
     K^t_C(x)=\min_p\{&|p|: C(p)\texttt{ halts with }x\\&\texttt{ in time }t(|x|)\}
 \end{align*}
 Intuitively, as $t(|x|)$ becomes larger the length of the program that needs this time becomes shorter; as it takes more time for them to create a large artifact than it takes an already larger program (we do not take into account the inner mechanisms of the programs which can be regarded as a shortcoming of this formulation). But is it always economical to sacrifice in time for a small gain in space? In Conway's game of life, a simple initial configuration can give rise to non-trivial organized complexity after running for exponential time which may lead us to think that this final configuration is valuable: it is attainable from its initial simplicity but only with much endeavours. However, if it can be reached from another slightly more complex initial settings within polynomial time, then it no longer seems as valuable as it did before. Therefore, a more error-free way of defining value is: an object is valuable only if all of its short programs are slow-running; in other words almost all of its algorithmic probability comes from slow and short programs. This is captured succinctly by the Logical depth $depth_c(x)$ \cite{a83} which we adopt as a formal measure of $x$'s value.
 $$depth_s(x)=\min_p\{time(p):C(p)=x,\;|p|\leq K_C(x)+s\}$$
 Here $s$ denotes the significance of the witness program $p$ in terms of its contribution to $x$'s existence: the less $s$ is, the more probable it is for $p$ to be $x$'s actual source on $C$; thus $p$'s computational effort is also the most probable way in which $x$ was created.
 
 \subsection{Compressibility$\not=$Value}
 Note that such definition of value also protects us from certain pitfalls; like claims that highly compressible and regular objects are valuable; rather we seem to attach value to objects that are meaningful and often such meaning is acquired over time. In Colton's Art exhibit example \cite{a108} a similar scenario is described where an art enthusiast deems a painting of random dots more valuable than another similar painting as it had a meaning attached to it provided by the painter: the random dots depicted his personal relationships and how he felt about them.
 
\begin{figure}[ht]
\centering
\begin{subfigure}{.5\linewidth}
  \centering
  \includegraphics[scale=0.20]{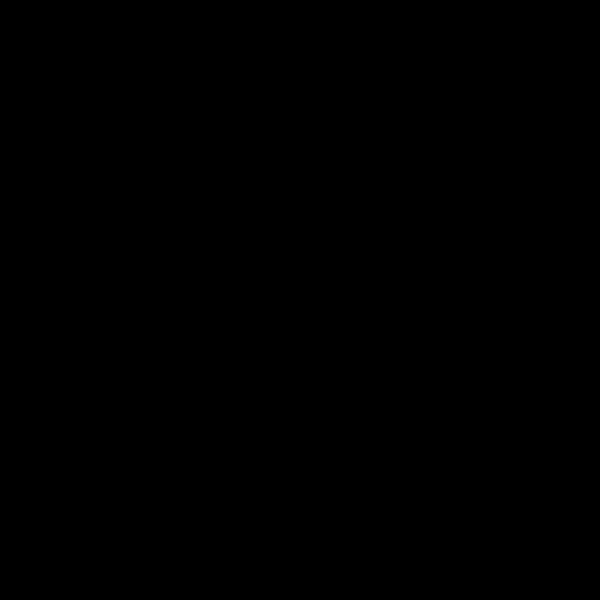}
  \caption{}
  \label{fig:subfirst}
\end{subfigure}%
\begin{subfigure}{.5\linewidth}
  \centering
  \includegraphics[scale=0.23]{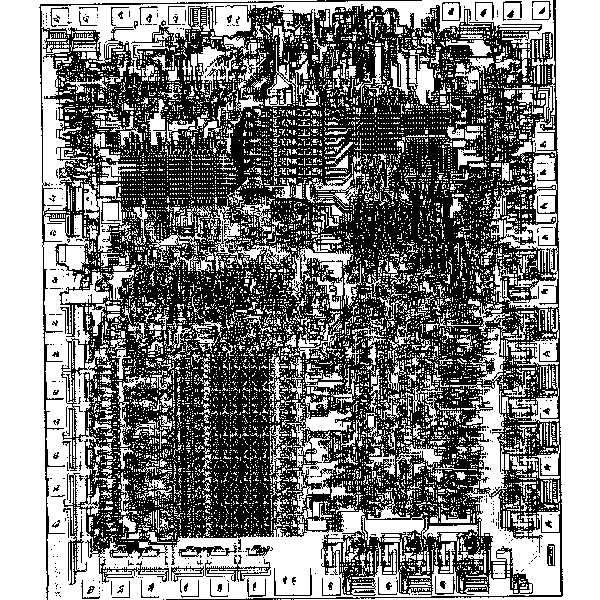}
  \caption{}
  \label{fig:sub2first}
\end{subfigure}
\caption{The mono-coloured image on the left is more compressible than the complex one on the right; yet the latter takes more decompression time from its short program. Image Courtesy: \cite{a124}}
\label{fig:test1}
\end{figure}

\begin{figure}[ht]
\centering
\begin{subfigure}{.5\linewidth}
  \centering
  \includegraphics[scale=0.20]{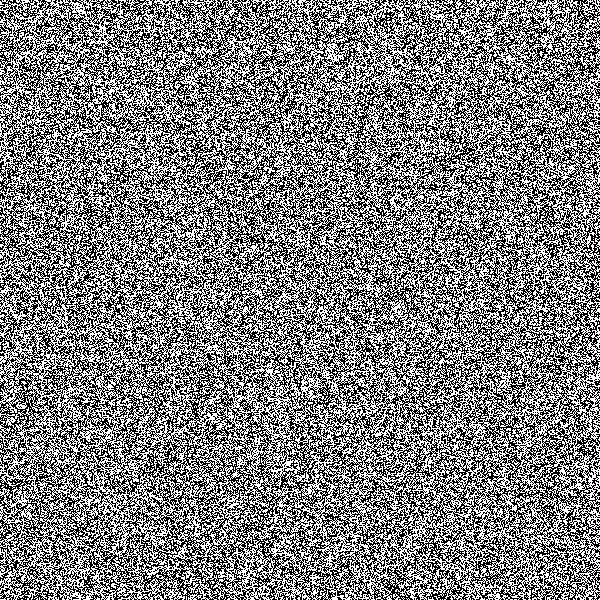}
  \caption{}
  \label{fig:sub1second}
\end{subfigure}%
\begin{subfigure}{.5\linewidth}
  \centering
  \includegraphics[scale=0.23]{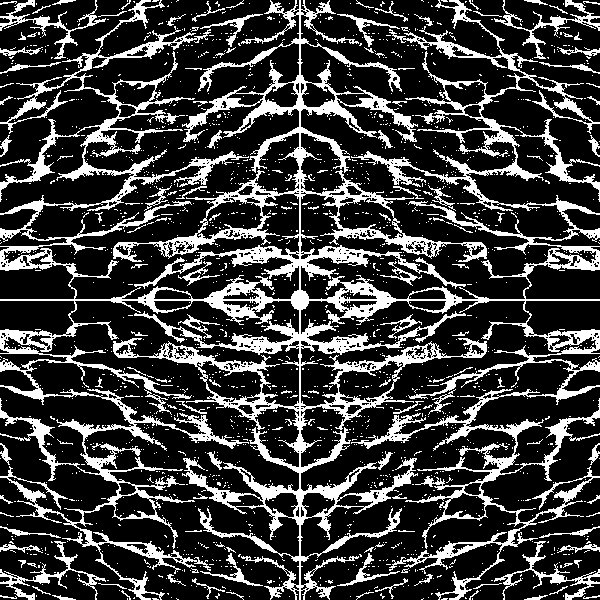}
  \caption{}
  \label{fig:sub2second}
\end{subfigure}
\caption{The image on the left is purely random noise. Hence a suitable program that generates it will contain a lot of verbatim description taking little time to reproduce the image. The image on the right by the virtue of being highly compressible takes more decompression time from its short program. Image Courtesy: \cite{a124}}
\label{fig:test2}
\end{figure}

 In an interesting study \cite{a124}, the authors showed that complex but non-random images that we would normally assign value to takes more time to generate from their lossless PNG (Portable network graphics) compression (Figure \ref{fig:test1}). 
On the other end of the spectrum, highly random objects which are not compressible and are often of little value to us, can be generated quickly from their similar-length programs (Figure \ref{fig:test2}).
 
 It is important to note here that this object-dependent notion of value does not consider the cultural context in which it was was conceived nor the observer's perspective which may differ from person to person. Instead, what it correctly predicts is that valuable objects are extremely rare: there are far more random objects in the universe than there are highly regular compressible ones; even less are among those which are programs and go through a long evolution (computation) to create something meaningful. This catapults us to the discussion of the computational complexity associated with value. 
 \section{Complexity of Creating Value}
 Consider creating an artifact $x$ with value $T$ (some large number) using the following diagonalization procedure \cite{a83}:  we enumerate all programs $p:|p|\leq T$, run each of them for time $T$, check whether the cumulative probability $\sum_{p:C(p)=x}2^{-|p|}$ of programs that halt with output $x$ is significant or larger than $2^{-|x|}$ (which means $x$ is compressible and has a more probable source than a verbatim description). If we do find an object $x$ that is less probable to be created by these $T$-fast programs, then we can say with high probability that $x$ has value at least $T$ (as programs taking more time than $T$ may be able to produce $x$; the algorithm we just described is certainly one among them) This is a very slow process taking more than $2^T$ to create an object with value $T$. However, this process can not be sped up due to the slow-growth law which states that a fast deterministic process can not turn a shallow or valueless object into a deep and valuable one. If it did then this process along with the shallow object (can be an empty string) could be encoded as a program that generates the valuable object quickly contradicting the fact that the object is valuable. This phenomenon is best illustrated by a great work of autobiography: the existence of such a literature would be impossible without the meaningfully lived life of the author.
 
 Given an artifact $x$ it is natural to think that starting from some initial long program (``print $x$'') we can incrementally update it by exploring shorter programs by running them until termination and check whether they output $x$. Unfortunately, since halting of a program is undecidable, even if we encounter a short program that eventually outputs $x$, without a preconceived idea of the object's value we would not know how long we should keep the program running. A halting sequence $H$ however, whose $i$th bit is 1 iff the $i$th program halts (we enumerate all possible programs with lexicographically increasing length) has the ability to speed up this process: with the knowledge that a particular program halts with certainty the only thing that remains to check is whether the program outputs $x$ \cite{a83}. By the slow growth law $H$ itself is logically deep and useful (in fact the time needed to generate this sequence from a small program as described above is unbounded)  as it aids the creation of all other valuable objects. An art appraiser with a vast knowledge of what constitutes valuable works, enjoys such efficiency when it comes to evaluating a new artwork; here we say their collective knowledge is useful which possibly took a long time to build. 
 
 Another natural question to ask is: whether there are ways of generating depth more closely approximating the maximum allowed by the slow growth law: depth $T$ in time
$T$. One way of doing so would be simply to generate a string of $T$ zeros. Then this string is valuable whenever $T$ itself is deep \cite{a83}. Time $O(T)$ is clearly both necessary and sufficient to
generate this string, but a string of zeroes is not useful and it would be more satisfying to find an example of an efficiently generated object deeper than its own bulk. It turns out this can only be possible if the creator has access to a random oracle with the help of which they can convert a simple string like $O^n$ to a deep string that takes polynomial time $O(kn)$ to be generated for some constant $k$; but it will take any normal observer proportional to $O(kn^2)$ time to generate the object from the same simple source \cite{a83}. The paintings of Jackson Pollock has such quality: at a first glance they seem random and producible by a trivial highly complex program. However, Pollock's drip paintings have been clearly identified as masterpieces which means they have overwhelming probability to be generated by very slow-running short programs \cite{a98,a126} regardless of the time taken by its creator.
 \section{Complexity of Verifying Value}
Consider Colton's art exhibition example again: the painter supplies the observer with a short program that simulates the events and relationships of the painter's life up to the point of creating the random dots painting after which the program terminates. If the program takes exponential time $O(2^{|x|})$ in the painting's length $|x|$ and the observer only has polynomial time, is it possible for a sceptical observer to be convinced of the meaning or value in the painting? Due to the deterministic nature of our setup, the answer points towards the negative: the provided program has to be run in order to see whether it outputs $x$ in the end. The exponential time-bound here is not too ambitious as typically, the most valuable or useful strings (like the Halting sequence $H$) take more than recursively bounded time to be generated \cite{a83}. 

Note that this problem gets much more difficult if the painter instead claimed the following: $p$ outputs $x$ in $O(2^{|x|})$ time and no other shorter program does so more efficiently. Now, the observer additionally has to enumerate all programs up-to length $|p|$, run them for $O(2^{|x|})$ and reject the claim if any of the programs halts within the time with output $x$. More difficult still is the claim that $x$ is logically deep with exponential time bound with significance at least $|p|-K_C(x)$ as it would require the observer to also generate all $|p|$ length programs to check whether $O(2^{|x|})$ is the minimum time taken by the programs to generate $|x|$. In general, the more valuable the object is, the harder it gets to detect such value. 

Because of the interactive nature of the value-verification problems let us see what an interactive proof (IP) system \cite{a127} might look like. To determine the value of $x$ we must first select a threshold runtime which can be $O(2^{|x|})$ as before. An all-powerful prover can find the shortest program $p$ that outputs $x$ in the designated time and to convince the verifier that no other same length or shorter program can output $x$ faster, they can provide a proof as following: the prover can run all programs with length $\leq|p|$ for $O(2^{|x|})$ time, collect the outputs of the halting programs and send it to the verifier. The problem then reduces to finding $x$ in this string. Note that, since the programs are allowed to run for $O(2^{|x|})$ time the largest output can have length at-most $O(2^{|x|})$ and in the worst case, the proof can have length $O(2^{|p|+|x|+O(1)})$. If the communication overhead between the two parties is allowed to be exponential then an exponential verifier can easily with high probability accept or reject the proof (each message from the prover contains a $O(2^{|x|})$ length output which the verifier compares with $|x|$ and there are at most $O(2^{|p|})$ messages). However, it is not clear whether a polynomial time-bounded verifier can accept or reject such proofs with high probability and if they have access to a random oracle (which is a standard addendum to a verifier in IP) what must its purpose or computational complexity be.  

Note that such austere difficulty in assessing value comes from the fact that compression is generally a more difficult task that decompression \cite{a128}. To find a non-trivial compression $q(x):C(q(x))=x\texttt{ and }q(x)<|x|$ the search space becomes exponential to find some encoding which decompressed to $|x|$. Besides some objects are cryptographically pseudo-random \cite{a83,a129}, meaning it is extremely difficult to pin-point a non-random source for them (highly sophisticated and complex artworks are good examples of this). While this maybe unsatisfactory, it might still be possible to efficiently detect a logically shallow or valueless artifact. A dishonest prover may claim that a shallow object is valuable but being shallow it is guaranteed to have at least one fast short program that generates the object. A clever observer after some bounded time analysis (or compression) of the object can detect such short program and refute the claim with high probability \cite{a128}.
\section{Conclusion and Future Direction}
We presented an object-dependent analysis of value rooting it in Algorithmic Information Theory and discussed the computational difficulty associated with determining and creating such value. Although such formulation does not take into account the object's cultural significance, its ability to be understood by viewers, or any other social properties, various authors \cite{a122,a133} have made similar connections between an object's value and its buried computational effort. This is also related to a notion of ``organized complexity'' found in nature: a priori the most probable explanation for the sequence of bases in a biological object like a DNA molecule is that it is the product of an extremely long evolutionary process \cite{b1}. 

We presented a time-dependent discussion but analogous results can be shown for space-bounded computation which illustrates the difficulty a mechanism with limited-memory faces when critiquing or evaluating a valuable object. Moreover, space-bounded computation sheds light on another exciting research direction: an object which stores information about its own creation. This is formalized with reversible computation where a computer along with the output supplies all the discarded information that can be used to trace back to the original program that is the source of the object. Then storing information from a long computation quickly becomes intractable, specially for one-way naturally occurring processes: a bottle of sterile nutrient solution has low complexity, but if inoculated with a single bacterium it can quickly turn into a bacterial culture. The rapid growth of bacteria following introduction of a seed bacterium is a irreversible process and even without the seed the forward process is vastly more probable than its reverse: transformation of bacteria into low-complexity energy nutrient. Complexity thus not to increase quickly, except with low probability, but can increase slowly over geological time, which makes taking snapshots of such progress extremely space-inefficient.

Another future direction for research will be to situate the intractable problems in value-determination within appropriate complexity classes. As we noted in the previous sections, it is not possible to reliably verify the value of an object in a classical settings within reasonable time-bound, however the quantum realm can lend a hand. One way to approach this is the observation that a valuable or logically deep object is one that is not in the set of shallow or valueless objects (our example of diagonal creation of value made use of this principle) and shallow or even weekly deep objects \cite{a83} are abundant in nature. This problem is remarkably similar to the Group non-membership problem \cite{a130} which belongs to the Quantum Merlin-Arthur (QMA) \cite{a132} complexity  class. We do not go into details but including or excluding the value problems from such classes will give a good idea about their true complexities and whether quantum techniques like entanglement and super-dense coding plays any role in efficient verification of value. 
\section{Acknowledgement}
I would like to thank Prof. Daniel Brown for his valuable comments and suggestions on the topics of this short paper.

\bibliographystyle{unsrt}  
\bibliography{references}  %%% Remove comment to use the external .bib file (using bibtex).

\begin{thebibliography}{10}

\bibitem{b1}
Ming Li and Paul~M.B. Vit{\'a}nyi.
\newblock {\em An Introduction to Kolmogorov Complexity and Its Applications}.
\newblock Springer Publishing Company, Incorporated, 4th edition, 2019.

\bibitem{a125}
R.~Landauer.
\newblock Irreversibility and heat generation in the computing process.
\newblock {\em IBM Journal of Research and Development}, 5(3):183--191, 1961.

\bibitem{a83}
C.~H. Bennett.
\newblock Logical depth and physical complexity.
\newblock In {\em A Half-Century Survey on The Universal Turing Machine}, page
  227–257, USA, 1988. Oxford University Press, Inc.

\bibitem{a108}
Simon Colton.
\newblock Creativity versus the perception of creativity in computational
  systems.
\newblock In {\em AAAI Spring Symposium: Creative Intelligent Systems}, pages
  14--20. AAAI, 2008.

\bibitem{a124}
Hector Zenil, Delahaye Jean-Paul, and Cédric Gaucherel.
\newblock Image characterization and classification by physical complexity.
\newblock {\em Complexity}, 17:26 -- 42, 01 2012.

\bibitem{a98}
Lu\'{\i}s Antunes and Lance Fortnow.
\newblock Sophistication revisited.
\newblock {\em Theor. Comp. Sys.}, 45(1):150–161, April 2009.

\bibitem{a126}
Tiasa Mondol and Daniel~G. Brown.
\newblock {\em To Appear in the Proceedings of the Twelfth International
  Conference on Computational Creativity, Mexico Virtual, September 14-18,
  2021}, 2021.

\bibitem{a127}
S~Goldwasser, S~Micali, and C~Rackoff.
\newblock The knowledge complexity of interactive proof-systems.
\newblock In {\em Proceedings of the Seventeenth Annual ACM Symposium on Theory
  of Computing}, STOC '85, page 291–304, New York, NY, USA, 1985. Association
  for Computing Machinery.

\bibitem{a128}
Stephen Fenner and Lance Fortnow.
\newblock Compression complexity, 2017.

\bibitem{a129}
Y.~Liu and R.~Pass.
\newblock On one-way functions and kolmogorov complexity.
\newblock In {\em 2020 IEEE 61st Annual Symposium on Foundations of Computer
  Science (FOCS)}, pages 1243--1254, Los Alamitos, CA, USA, nov 2020. IEEE
  Computer Society.

\bibitem{a122}
Cl{\'e}ment Vidal and Jean-Paul Delahaye.
\newblock Universal ethics: Organized complexity as an intrinsic value.
\newblock In Georgi~Yordanov Georgiev, John~M. Smart, Claudio~L.
  Flores~Martinez, and Michael~E. Price, editors, {\em Evolution, Development
  and Complexity}, pages 135--154, Cham, 2019. Springer International
  Publishing.

\bibitem{a133}
Seth Lloyd and Heinz Pagels.
\newblock Complexity as thermodynamic depth.
\newblock {\em Annals of Physics}, 188(1):186--213, 1988.

\bibitem{a130}
Lov~K. Grover.
\newblock A fast quantum mechanical algorithm for database search.
\newblock In {\em Proceedings of the Twenty-Eighth Annual ACM Symposium on
  Theory of Computing}, STOC '96, page 212–219, New York, NY, USA, 1996.
  Association for Computing Machinery.

\bibitem{a132}
C.~Marriott and J.~Watrous.
\newblock Quantum arthur-merlin games.
\newblock In {\em Proceedings. 19th IEEE Annual Conference on Computational
  Complexity}, pages 275--285, Los Alamitos, CA, USA, jun 2004. IEEE Computer
  Society.

\end{thebibliography}
%%% and comment out the ``thebibliography'' section.

%%% Comment out this section when you \bibliography{references} is enabled.

\end{document}